\documentclass[conference]{IEEEtran}
\ifCLASSINFOpdf
\else
\fi
\usepackage{amsmath}
\usepackage{amssymb}
\usepackage{calrsfs}
\newtheorem{Definition}{Definition}
\newtheorem{lemma}{Lemma}

\newtheorem{proposition}{Proposition}
\newtheorem{corrollary}{Corrollary}

\usepackage{graphicx}
\graphicspath{%
    {converted_graphics/}
    {Z:/mydocs/BORKARpub/WiOPT2016/WiOPT_figures/}
    {Z:/mydocs/BORKARpub/WiOPT2016/ArxivSubmit/Arxiv_Figures/}
}
\begin{document}
%
\title{Optimal Spread in Network Consensus Models}

\author{\IEEEauthorblockN{F. Y. Hunt}
\IEEEauthorblockA{Information Technology Laboratory, Mail Stop 8910 \\
National Institute of Standards and Technology\\
Gaithersburg, Maryland, 20899\\
Email: fern.hunt@nist.gov}}


%


\maketitle

\begin{abstract}
In a model of network communication based on a random walk in an undirected graph, what subset of nodes (subject to constraints on the set size), enable the fastest spread of information? The dynamics of spread is described by a process dual to the movement from informed to uninformed nodes. In this setting, an optimal set $A$ minimizes the sum of the expected first hitting times $F(A)$, of random walks that start at nodes outside the set. 

In this paper, the problem is reformulated so that the search for solutions to the problem is restricted to a class of optimal and "near" optimal subsets of the graph. We introduce a submodular, non-decreasing rank function $\rho$, that permits some comparison between the solution obtained by the classical greedy algorithm and one obtained by our methods. The supermodularity and non-increasing properties of $F$ are used to show that the rank of our solution is at least $(1-\frac{1}{e})$  times the rank of the optimal set. When the solution has a higher rank than the greedy solution this constant can be improved to $(1-\frac{1}{e})(1+\chi)$ where $\chi >0$ is determined a posteriori.
\\
\end{abstract}


%
\IEEEpeerreviewmaketitle

\section{Introduction} \label{S:intro}
The study of information spread (or dually consensus) in complex networks has been the subject of intense research in the past decade for example \cite{Olfati}, \cite{Clark}, \cite {Rao}, \cite{Kempe}, \cite{Richardson} where the role  of  distinguished subsets of nodes such as "leaders"  in consensus models and "influential spreaders" in models of information spread  is studied. In particular the research reported in references \cite{Richardson}, \cite{Kempe}, \cite{Borgs} have developed methods for obtaining optimal spreaders –as determined by some measure of subset performance. Another substantial body of related work is concerned with the construction and performance analysis of algorithms for efficient information spread, for example the so-called push/pull algorithms \cite{Giakkoupis}, the independent cascade model \cite{Borgs}, a random averaging scheme \cite{Boyd} and the GOSSIP model of \cite{Censor-Hillel}. In this paper, our focus will be on the first issue: the identification of optimal spreaders in a network.  We will use a random walk communication model and an objective function associated with this process. Results of this research are relevant to the design of algorithms for routing in wireless communication systems when location information is not available \cite{Rao,Jadbabaie}, identification of influential individuals in a social network \cite{Kempe} and in sensor placements for efficiently detecting intrusions in computer networks \cite{Krause}. 

   Given  a connected graph  $G=(V,E)$ with $N$ vertices $V$ and edges $E$, information spreads through the network by a process that is dual to the direction of the random walk (see \cite{Barahona}). An optimal spreader in our setting is defined in terms of a set function $F$ where for a subset $A \subset V$, $F(A)$ is the sum of mean first arrival times to $A$ by random walkers that start at nodes outside of $A$. If $A$ is an effective target set for the random walks  (dually an effective spreader) then $F(A)$ is small. Thus the optimal set (subject to a cardinality constraint $K$)  minimizes $F(A)$ subject to $|A| \leq K$,
\begin{equation}\label{E:opt}
\min_{ A \subset V,\\\ |A|\leq K} F(A).
\end{equation}
Recall that a random walker situated at a node $i \in V$, moves to a neighboring node $j \in V$ in a single discrete time
step with probability, 
%
\begin{equation}
\setlength{\nulldelimiterspace}{0pt}
Prob(i,j)=\left\{\begin{IEEEeqnarraybox}[\relax][c]{l's}
p(i,j), &for $(i,j) \in E$\\
0  &otherwise
\end{IEEEeqnarraybox}\right.
\end{equation}

{\bf NOTE:} In this discussion $p(i,j)=1/deg(i)$ where $deg(i)$ is the degree of node $i$. However any probabilities for which
the resulting Markov chain is ergodic can be used. \\

The matrix $\mathcal{P}=(p_{ij})_{i,j=1 \cdots N}$ is the transition matrix of a Markov chain which in
this paper, is assumed to be irreducible and aperiodic (\cite{Kemeny}).   Starting at node $i \notin A$, a random walker first reaches the set $A$ at a hitting time $T_{A}=\min\{n >0: X_{n} \in A \}$, where $X_n$ is the node occupied by the walker
at time $n$. Denoting the expected value of this time by $h(i,A)=\mathcal{E}_{i}[T_{A}]$, the value of $F$ at $A$ is expressed as
%
%
\begin{equation} \label{E:hitnumber}
F(A)={\sum}_{i \notin A}h(i,A).
\end{equation}

Given $A$, $F(A)$ can be evaluated by solving a suitable linear equation. Indeed
a standard result in Markov chain theory \cite{Kemeny} tells us that $h(i,A)$ is the ith component of the vector $\mathsf{H}$,
which is the solution of the linear equation,
%
\begin{equation}\label{E:lineq}
{\LARGE\mathsf{H}}=\bf{1}+\mathcal{P}_{A}{\LARGE\mathsf{H}}
\end{equation}
where $\mathbf{1}$ is a column vector of $N-|A|$ ones and  $\mathcal{P}_{A}$ is the matrix that results from
crossing out the rows and columns of $\mathcal{P}$ corresponding to the nodes of $A$.

 Borkar, Nair and Sanketh \cite{Borkar}, introduced the optimization problem (\ref{E:opt}) and showed that for subsets
$A\subseteq B \subseteq V \,$  and $ j \in V$, $F(A)-F(A\cup \{j\})\geq F(B)-F(B\cup\{j\})$, that is, $F$ is a supermodular function. Thus $-F$ is submodular and when bounded our problem is an instance of submodular maximization, a classic problem in combinatorial optimization. In 1987, Nemhauser, Wolsey and Fisher \cite{NemWolFish} showed for a bounded submodular function that a set constructed by the greedy algorithm has an approximation ratio of $(1-1/e)$. More recently, Borgs, Brautbar, Chayes and Lucier \cite{Borgs} and  Sviridenko,Vondrak and Ward in \cite{SvirVonWard}, showed that approximations of comparable quality could be obtained very efficiently using different methods.
 To minimize the convergence rate to consensus of a leader-follower network, Clark, Bushnell and Poovendran \cite{Clark}
 considered a supermodular function closely related to ours and showed that the greedy algorithm produces an approximation that is within $(1-1/e)$ of optimal.

In this paper we will discuss a method that obtains an exact or approximate solution to (\ref{E:opt}) by
introducing additional constraints in the problem that are based on properties of
the underlying graph. Observing that a vertex cover of the graph with $C$ vertices
is an optimal set for $K = C$, sets of cardinality $C$ or less can be assigned a ranking relative to it.
Using the rank (introduced in section \ref{S:optnopt}), we define a class of optimal and near optimal sets $\mathnormal{L}_{\nu,C}$,
where $\nu$ is the minimum rank of sets in the class. Here we consider $\nu$ as a measure of the quality of the approximation. 
To solve the problem for $K < C$, we choose a collection of sets $\mathbf{S}\subset \mathnormal{L}_{\nu,C}$. Each set
in $\mathbf{S}$ has cardinality $m$-- where $m$ is the minimum cardinality of sets in $\mathnormal{L}_{\nu,C}$. Note that 
the exact solution is in $\mathnormal{L}_{\nu,C}$ if $m< K < C$.
The output of this method is the best set that results from a greedy extension of each
set in $\mathbf{S}$, to a set of cardinality $K$. The method requires the determination of sets of cardinality $m$ each of pre-determined quality $\nu$ and the computational effort involved as discussed in section \ref{S:effort} is $O(N^{m+3})$. We assume that $m \ll K$ so a natural question is given $m$ what quality $\nu$ can be expected? Conversely given a required solution quality $\nu$, what $m$ is needed? 

The plan of the paper is as follows: Section \ref{S:newsolution} contains a definition and discussion of optimal and near optimal sets ranked relative to a vertex cover of the graph $G$ of cardinality $C$. We demonstrate how the method is applied to a graph using a collection of near optimal sets that are subsets of the vertex cover in section \ref{S:optnopt}. If every vertex cover contained optimal sets as subsets, it would make sense to use this choice consistently. Unfortunately, optimality of a set is generally not preserved by the addition or deletion of elements, otherwise the greedy algorithm would always yield exact solutions.
We remedy this situation in part by selecting a group $\mathbf{S}$ of $m$ element sets in $L_{\nu,,C}$ that contain a class of subsets satisfying the axioms of a greedoid (\cite{Korte} and see Section \ref{S:closure}). Its feasible sets are closed under the addition and deletion of certain elements. Moreover all feasible sets of cardinality $n >m$ are in $L_{\nu, C}$ and are therefore optimal or near optimal. In general, the greedoid is not unique and it may or my not contain optimal sets of required cardinality $K$. However any offered solution of our method that is feasible will be near optimal with some pre-defined quality. Sufficient conditions for the existence of $\mathbf{S}$ are stated in Section \ref{S:closure} and the details of the greedoid construction can be 
found in \cite{Hunt}. 
We also demonstrate the method on a second graph where $\mathbf{S}$  is chosen to be a group of feasible sets of a greedoid. 
 In Section \ref{S:quality}, the quality of the approximation is evaluated in terms of the ranking function $\bar{\rho}$ introduced 
in section \ref{S:optnopt} After normalizing $F$, we obtain $\rho$, a bounded submodular set function with $\rho(\emptyset)=0$. We can apply the results in \cite{NemWolFish}, 
to show that the ratio of the rank of our approximation to that of the optimal set is at least $(1-\frac{1}{e})$. Moreover,
the approximation can be compared to the other solutions obtained by the greedy extension of sets of cardinality less than $m$ including the classic greedy method that starts with a one element set. In particular, if the rank of a greedy solution is less than $\nu$, then the solution $S^{*}$ obtained by our method satisfies an inequality  that improves the $(1-\frac{1}{e})$ bound,
\begin{equation} \label{E:a-posteriori}
\rho(S^{*})\geq (1+\mathcal{\chi})(1-\frac{1}{e}) \rho(\mathcal{O}_{K})
\end{equation}
where $\mathcal{\chi}>0$ is a constant determined \textit{a posteriori} and $\mathcal{O}_{K}$ is an optimal solution of equation (\ref{E:opt}).

\section{Finding and Approximating Optimal Sets} \label{S:newsolution}
\subsection{Maximal Matches}\label{S:maxmatch}

The optimization problem as posed in equation(\ref{E:opt}) assumes no advance knowledge about the optimal set or any
other possibly related sets. We first consider a process of obtaining optimal sets by using
subsets of existing ones.
Let $A$ be a vertex cover (not necessarily a minimal one). Since every edge is incident to an element of $A$, a random walker starting at a vertex $i$ outside of $A$ must hit $A$ at the first step. That is $h(i,A)=1$. Now equation (\ref{E:lineq}) implies that $h(i,A)\geq 1$ so it follows that $A$ must be an optimal set for its own cardinality. Thus a solution for $C=|A|$ is obtained by constructing a vertex cover. Fortunately a maximal match can be constructed by a simple greedy algorithm and its vertices are a vertex cover with cardinality $C \leq 2*(cardinality\ of \ a \ minimum \ vertex \ cover)$ \cite{Cormen}. Therefore without loss of generality we turn our attention to the solution of problem (\ref{E:opt}) for $K \leq C$.

\subsection{Optimal and Near Optimal Sets}\label{S:optnopt}
We introduced a measure of the spread effectiveness of sets in Section \ref{S:intro}, equation (\ref{E:hitnumber}). It will be convenient to
convert this to a rank defined on subsets of $V$. In particular, suppose there exists a vertex cover with $C$ vertices.
We will order all non-empty subsets $A \subseteq V$ such that $|A|\leq C$ with a ranking function $\bar{\rho}(A)$ defined as,
\begin{equation}\label {E:rank}
\bar{\rho}(A)=\frac{F_{max}-F(A)}{F_{max}-F_{min}} 
\end{equation}
where $F_{max}=\max_{\emptyset \ne A\subseteq V, |A|\leq C}\large F(A)$, and $F_{min}$ is the corresponding minimum. $F_{min}$ can be calculated by computing $F$ for a maximal match of cardinality $C$, while $F_{max}$ is the maximal value of $F$ among all one element subsets. We assume that $F_{max}\ne F_{min}$. If this were not the case, $F(A)$ would be have the same value for any non-empty subset $A$ with $|A| \leq C$. Thus any $A$ would be a solution of the problem.

\begin{figure}[tbp] 
  \centering
  \includegraphics[width=2.5in,height=1.61in,keepaspectratio] {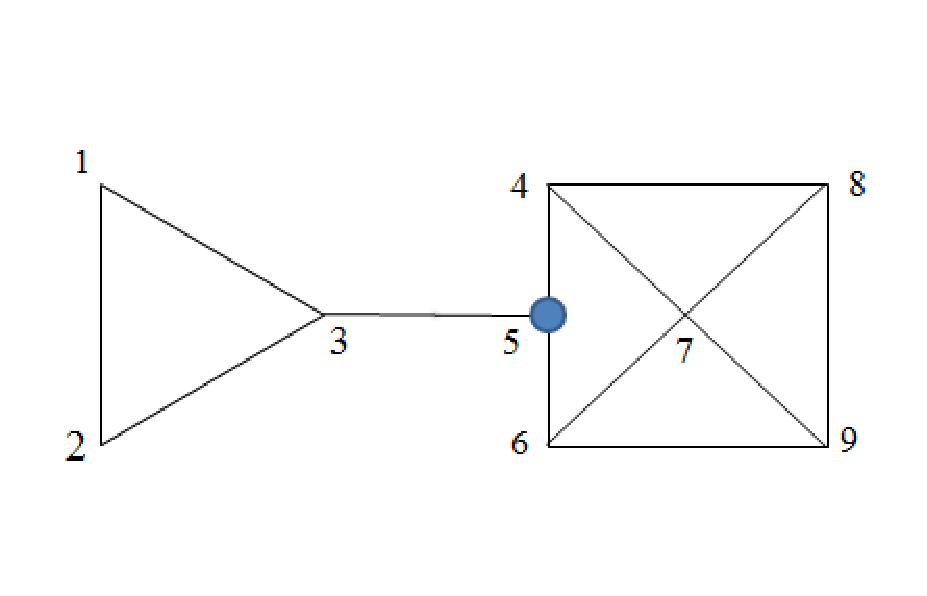}
  \caption{Graph with N=9 vertices, shows optimal set for K=1 (colored)}
  \label{fig:JW8K1}
\end{figure}

\begin{figure}[tbp] 
  \centering
  \includegraphics[width=2.56in,height=0.804in,keepaspectratio]{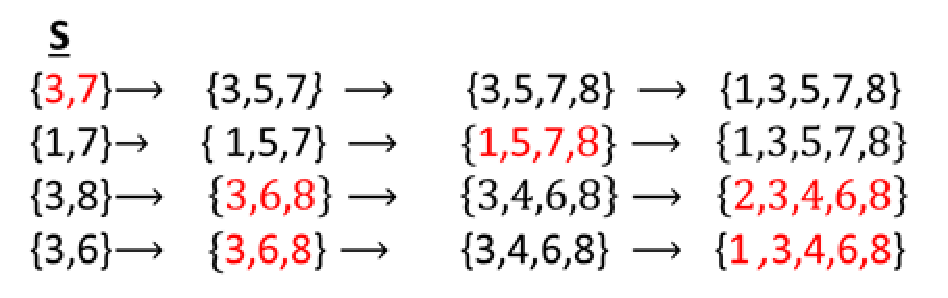}
  \caption{Optimal Sets for K=4,5 obtained by greedy extension of $\mathbf{S}$}
  \label{fig:VC_Method}
\end{figure}

If $A$ is optimal and $|A|=C$ then $\bar{\rho}(A)=1$ conversely the the worst performing set has value $0$.
Thus for a constant $\nu \, ( 0<\nu \leq 1$) and $C$, the non-empty set
\begin{equation}\label{E:Lck}
\mathnormal{L}_{\nu,C}=\{A: A\subseteq V, |A|\leq C , \bar{\rho}(A)\geq \nu\}
\end{equation}
 defines a set of optimal and near optimal subsets, with the degree of near optimality depending of course on $\nu$. 
Let $m$ be the smallest cardinality of sets in $\mathnormal{L}_{\nu,C}$. 
Starting with a collection of sets $\mathbf{S} \subset \mathnormal{L}_{\nu,C}$ of size $m$, our method is to seek a solution 
to problem (\ref{E:opt}) by greedily augmenting each set until it reaches the desired size $K$. The offered approximation is the best (has the lowest $F$ value) of these extended sets. We can always find a $\nu$ and $C$ so that $\mathnormal{L}_{\nu,C}$ contains the optimal set of cardinality $K$ but we do not have a proof that the approximation generated by subsets of a vertex cover is optimal. However since our solution is a superset of sets in $\mathnormal{L}_{\nu,C}$, it is also in $\mathnormal{L}_{\nu,C}$ and therefore has minimum rank $\nu$. We illustrate the method with an example. Figure \ref{fig:JW8K1} shows a graph with $N=9$ vertices along with the vertices of optimal sets for $K=1$.  To solve the problem for $K=4$, we note that the class of optimal and near optimal sets based on $C=8$ and $\nu=.90$ has minimum set size $m=2$. The set $\mathcal{M}=\{1,3,5,6,7,8\}$  is a vertex cover (calculated from the maximal match algorithm). We define $\mathbf{S}$ to be the two element subsets of $\mathcal{M}$ that are 
in $\mathnormal{L}_{.90,8}$. The first column of Figure \ref{fig:VC_Method} lists these sets and subsequent columns show the results of one element extensions of $\mathbf{S}$ until $K=5$. Optimal sets are shown in red.  In this example the offered approximation is optimal. This is also the case for extensions up to $K=5$. In this case we see that the method identifies optimal sets that are subsets of $\mathcal{M}$ as well as others that are not e.g. $\{2,3,4,6,8\}$, underlining the fact the method finds sets that are reachable by greedy extension of subsets of $\mathcal{M}$. The offered approximation for this method is guaranteed to be in 
$\mathnormal{L}_{.90,8}$. This is a consequence of Proposition \ref{P:upextend} which is discussed and proved in 
Section \ref{S:closure}\\
%


\section{Closure Property of Optimal and Near Optimal Sets}\label{S:closure}
\begin{figure}[tbp] 
  \centering
  \includegraphics[width=2.56in,height=2.08in,keepaspectratio]{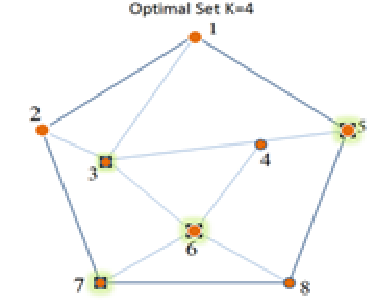}
  \caption{Graph with N=8, vertices. Vertices of optimal set K=4 shown as squares}
  \label{fig:BiggsK4B-vs3}
\end{figure}

\begin{figure}[tbp] 
  \centering
  \includegraphics[width=2.56in,height=1.23in,keepaspectratio]{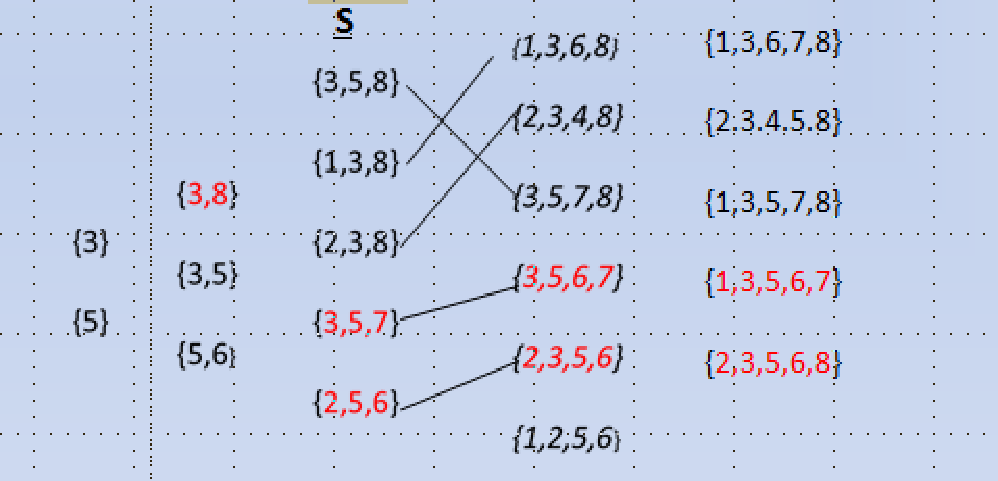}
  \caption{Greedoid constructed from optimal and near optimal sets $\mathnormal{L}_{.85,7}$ of graph in Fig \ref{fig:BiggsK4B-vs3}. Empty set not shown.}
  \label{fig:BiggsGREEDOID}
\end{figure}

In section \ref{S:optnopt}, we demonstrated our method of approximating a solution of problem  (\ref{E:opt}) based on greedy extensions of subsets of a vertex cover that are optimal or near optimal. Unfortunately a vertex cover can fail to have such subsets other than the vertex cover itself (see an example in \cite{Hunt}). This is the motivation for finding other classes of optimal and near optimal sets that permit the addition and deletion of elements. The structure we seek is conveniently described in terms of a generalization of the matroid known as a {\it greedoid}  \cite{Korte,Bjorner}.
\begin{Definition}\label{D:greedoid}
Let $\mathbf{E}$ be a set  and let $\mathcal{ F}$ be a collection of subsets of $\mathbf
{E}$. The pair $(\mathbf{E}, \mathcal{F})$  is called a \underline {greedoid} if $\mathcal{F}$ satisfies
\begin{itemize}
  \item $\mathbf{G1}:$ $\emptyset\in \mathcal{F}$ \label{null}
  \item $\mathbf{G2}:$ For $A \in \mathcal{F}$ non-empty, there exists an $a \in A$ such that $A \setminus \{a\} \in \mathcal{F}$ \label{access}
  \item $\mathbf{G3}:$ Given $X$, $Y$ $\in \mathcal{F}$ with $|X| > |Y|$, there exists an $x \in X\setminus Y$, such that $Y \cup \{x\} \in \mathcal{F}$ \label{augment}
\end{itemize}
\end{Definition}
A set in $\mathcal{F}$ is called \underline {feasible.} Note that $\mathbf{G2}$ implies that a  single element can be removed from a feasible set $X$ so that the reduced set is still feasible. By repeating this process the empty set eventually is reached. Conversely
starting from the empty set, $X$ can be built up in steps using the $\mathbf{G3}$ property.\\
We now show that $\mathnormal{L_{c,K}}$ satisfies condition $\mathbf{G3}$ of the definition
for any  $0 < c \leq 1$,\, $0\leq K \leq N$ (Proposition \ref{P:upextend}). The proof depends on the following lemma and uses
an adaptation of an argument in Clark et al \cite{Clark}
\begin{lemma}\label{L:mono}
Let $S \subseteq V$, $u \in V\setminus S$. Then $F(S) \geq F(S\cup \{u\})$. 
\end{lemma}
{\bf Proof:} Suppose $S$, a set of nodes is a target set for the random walk. Let $E_{ij}^{l}(S)$ be the event,
$E_{ij}^{l}(S)=\{ X_{0}=i \in V,\   X_{l}=j \in V\setminus S,\ X_{r} \notin S,\ 0 \leq r \leq l \}$.
Thus paths of the random walk in this event start at $i$ and arrive at $j$ without visiting $S$ during the interval $[ 0 , l ]$. 
Also define the event $F_{ij}^{l}(S,u)=E_{ij}^{l}(S) \cap \bigcup_{m=0}^{l} \{ X(m)=u\}$ where $u \notin S$. 
Paths in this event also start at $i$ and arrive at $j$ without visiting $S$, but  must visit the element $u$ at some time during the interval $[ 0 , l ]$. Since a path either visits $u$ in the time interval $[ 0 , l ]$ or it does not, it follows that:
\begin{equation}\label{E:seteq}
E_{ij}^{l}(S)=E_{ij}^{l}(S \cup \{u \})\cup F_{ij}^{l}(S,u)
\end{equation}
We have  $E_{ij}^{l}(S \cup \{u\}) \bigcap F_{ij}^{l}(S,u)=\emptyset$. This implies that,
\begin{equation}\label{E:indeq}
\chi(E_{ij}^{l}(S))=\chi(E_{ij}^{l}(S \cup \{u\}))+\chi(F_{ij}^{l}(S,u))
\end{equation}
and therefore:
\begin{equation}\label{E:notindeq}
\chi(E_{ij}^{l}(S)) \geq \chi(E_{ij}^{l}(S \cup \{u\})
\end{equation}
Here $\chi(A)$ is the indicator function of the set $A$. Recalling that $T_{S}$ is the hitting time for
set $S$, the following relation comes from taking the expection of $\chi(E_{ij}^{l}(S))$ on the left hand side of (\ref{E:notindeq}) summing over all $j \in V\setminus S$. Here $\mathcal{E}$ denotes expectation.
\begin{equation}\label{E:hitS}
\mathbf{Prob}\{ T_{S} > l | X_{0}=i\}=\mathcal{E}\left(\sum_{j \in V \setminus S}\chi(E_{ij}^{l}(S))\right)
\end{equation}
A similar result is obtained for $T_{S\cup \{u\}}$ from taking the expectation of $\chi(E_{ij}^{l}(S\cup\{u\}))$ on the right
hand side of (\ref{E:notindeq}) and summing over $j \in V\setminus S$.
Summing once again over all $l \geq 1$ results in the inequality,
\begin{equation}\label{E:hitsumineq}
h(i,S) \geq h(i,S\cup \{u\})
\end{equation}
\begin{proposition}\label{P:upextend}
For $0 < c\leq 1$ and $0 < K \leq N$, let $\mathnormal{L}_{c,K}$ be the class of sets defined in equation (\ref{E:Lck}).
Then $\mathnormal{L}_{c,K}$ satisfies condition  $\mathbf{G3}$.
\end{proposition}
{\bf Proof:} The conclusion follows from the definition of $\mathnormal{L}_{c,K}$ and the fact that $F$ is non-increasing.
$\Box$  \\ \\
Proposition \ref{P:upextend} establishes that $L_{c,K}$ satisfies the $\mathbf{G3}$ property for greedoids. However, $\mathbf{G2}$ does not hold. For example
if the set $A$ has cardinality $m$ where $m$ is the size of the smallest set in $L_{c,K}$ then $A\setminus \{a\}$ cannot be in 
$\mathnormal{L_{c,K}}$  for any element $a \in A$. Conversely, let $c_{n}=\max_{|X| \leq n}\rho(X)$. If $c_{m}\geq c>c_{m-1}$ then $m$ is the size of the smallest set in $\mathnormal{L_{c,K}}$. Define $G_n$ to be all sets in $\mathnormal{L}_{c,K}$ of cardinality $n$. To create a class of sets with the $\mathbf{G2}$ property, one constructs subsets of $G_m$ of size $n \leq m$ that  are "augmentable" i.e. that satisfy $\mathbf{G3}$, while sets $G_n$ for $n > m$ are culled so the remaining sets are supersets of the "augmentable" sets and therefore satisfy $\mathbf{G2}$. The greedoid will then consist of selected subsets and supersets of $G_m$. Conditions for the existence of "augmentable" subsets of $G_m$ and proof of the validity of the resulting greedoid construction can be found in
 \cite{Hunt}. Rather than repeat the details of these arguments here, we close this section with an example showing the greedoid of a graph (Figure \ref{fig:BiggsK4B-vs3}) and its use in the solution of (\ref{E:opt}).The minimum cardinality of a set in the class of optimal and near optimal sets $\mathnormal{L}_{.85,7}$ is $m=3$. These sets are used to create the greedoid depicted in Figure \ref{fig:BiggsGREEDOID}. Note that $\mathbf{G1}$-$\mathbf{G3}$ are satisfied.
Assume the optimal set for $K=4$ is unknown. Then our method in this case is to take $\mathbf{S}$ to be the three element sets in 
$\mathnormal{L}_{.85,7}$  that are feasible sets of the greedoid and perform a greedy extension of each set. In figure
 \ref{fig:BiggsGREEDOID} a line is drawn between a set and its greedy extension. We have also drawn greedy extensions of sets of cardinality $n < m$ as well.
The optimal sets are shown in red and so they are in the greedoid. The offered approximations are in fact exact. 

\section{Quality of the approximation}\label{S:quality}
\subsection{Comparison between the optimal solution and greedy solution} \label{S:submodcompare}

Following Ilev (\cite{Ilev}), $F$ can be defined for the empty set as 
\begin{equation}
0\leq F(\emptyset)=\max_{X \cap Y=\emptyset , X, Y \subseteq V} F(X) +F(Y)-F(X \cup Y) < \infty
\end{equation}
Thus by the definition of $\bar{\rho} \, \,$,  $\bar{\rho}(\emptyset)=\frac{F_{max}-F(\emptyset)}{F_{max}-F_{min}}$.
This means the normalized function defined on sets $A , \,$ $\rho(A)=\bar{\rho}(A)-\bar{\rho}(\emptyset)$ is bounded, submodular, non-decreasing. For the empty set we have $\rho(\emptyset)=0$.

Our offered solution is the result of a greedy extension of a group of $m$ element sets $\mathbf{S}$.  Using $\rho$ it can be compared to an $m$ element set that is the result of greedily adding single elements $m$ times. Call this set $S_{g}$. We first suppose that 
$S_{g}\in \mathbf{S}$. 
\begin{lemma} \label{L:NWF} Suppose $S_{g} \in \mathbf{S} \subseteq \mathnormal{L}_{\nu,C}$. Let $S_{g}^{(K)}$ be the K element set obtained from the greedy extension of $S_{g}$.  If $S^{*}$ is the offered solution, then
\begin{equation}
F(S^{*}) \leq F(S_{g}^{(K)})
\end{equation}
\end{lemma}
{\bf Proof:} $F(S^{*})$ is the minimum value of all the values obtained by the greedy $K-m$ extension of elements in $\mathbf{S}$.
$\Box$ \\
The set $S_{g}^{(K)}$ is also the result of greedily adding single elements $K$ times. Thus we may use \cite{NemWolFish} (section 4) and the definition of $\rho$ to conclude that
\begin{corrollary} \label{C:bound}
If $S^{*}$ is the solution constructed by the method described in sections \ref{S:optnopt} and \ref{S:closure}, then 
\begin{equation}
\rho(S^{*})\geq (1-\frac{1}{e})\rho(\mathcal{O}_{K}^{*})
\end{equation}
where $\mathcal{O}^{*}_{K}$ is the optimal solution of problem (\ref{E:opt}).
\end{corrollary}
Once $F(S^{*})$ and $F(S^{(K)}_{g})$ have been computed we can determine $\chi$ such that 
$\rho(S^{*})=(1+\chi)\rho(S^{(K)}_{g})$. Therefore if $F(S^{*})< F(S^{(K)}_{g})$ the bound in Corrollary \ref{C:bound}, can be strengthened.\\
\begin{proposition}  \label{P:betterbound} When $F(S^{*})<F(S^{(K)}_{g})$, so $\chi >0$, then
\begin{equation}
\rho(S^{*})\geq  (1+\chi)(1-\frac{1}{e})\rho(\mathcal{O}^{*}_{K}) \\
\end{equation}
\end{proposition}
If $S_{g} \notin \mathbf{S}$, the conclusion of Proposition \ref{P:betterbound} is still valid when the greedy extension to
a $K$ element set $S_{g}^{(K)}$ satisfies $\rho(S_{g}^{(K)}) < \nu$. Indeed by the closure property of $\mathnormal{L}_{\nu,C}$ (Proposition \ref{P:upextend}), $\rho(S^{*})\geq \nu$ and thus $\chi >0$. This bound is also valid for solutions obtained  using the greedy extension of sets of cardinality less than $m$ for which lower bounds of the type $(1-\frac{1}{e})$ have been established.
A lower bound of $(1-\frac{1}{e})$ was previously established by Borkar et al in \cite{Borkar} for $F$.
Specifically they proved there a lower  bound on the ratio of $F(S_g)-F(\{a\})$ 
to $F(\mathcal{O}^{*}_{K})-F(\{a\})$ where $S_g$ is the result of the greedy algorithm starting with singleton $a$.
%
\subsection{Computational effort and tradeoff with quality} \label{S:effort}
A rough estimate of the complexity of the method follows from realizing that the collection $\mathbf{S} \in$ 
$\mathnormal{L}_{\nu,C}$, has at most $\binom{N}{m},\,$  $m$ element sets. To determine whether or not a particular set is near optimal,
equation(\ref{E:lineq}) must be solved and this involves $O(N^{3})$ operations. Thus $\mathbf{S}$ is determined in 
$O(N^{m+3})$ operations. The greedy extension of an $m$ element to a $K$ element set involves $O((K-m)(N-m))=O(N^{2}))$ operations so that for the extension of every set in $\mathbf{S}$ we need $O(N^{m+2})$ operations. Overall then, the method requires $O(N^{m+3})$ operations.
It is desirable therefore to make $m$ as small as possible for example with $m \ll K$. However the size of $m$ affects the accuracy.
 Taking $\nu$ to be a measure of the quality of the approximation, we want to know given $m$, what $\nu$ can be expected?
 Conversely given a desired quality $\nu$, what $m$ is required?
 We will employ the elemental curvature of the rank function (see equation (\ref{E:rank})). Elemental curvature was used by Wang, Moran, Wang, and Pan \cite{Wang} in their treatment of the problem of maximizing a monotone non-decreasing submodular function subject to a matroid constraint. Recall from section \ref{S:submodcompare},  that $\rho$ is a  submodular, monotone and non-decreasing set function that vanishes on the empty set.

The elemental curvature of $\rho$ is defined 
over $\mathnormal{L}_{\nu,C}$ in terms of the marginal increase in the rank of a set when a single element is added to it.
First let $A$ be a set and $i \notin A$,
\begin{equation} \label{E:rankincr} 
\rho_i(A)=\rho(A \cup i)-\rho(A).
\end{equation}
Then for a fixed $A \in \mathnormal{L}_{\nu,C}$ set,
\begin{equation} \label{E:curveS}
k_{ij}(A)=\frac{\rho_{i}(A \cup j)}{\rho_{i}(A)}. 
\end{equation}
The curvature is defined then as,
\begin{equation} \label{E:curvature}
\kappa=\max\{ k_{ij}(A) : A \subset \mathnormal{L}_{\nu,C}\,\, , i\ne j,\, i, j \notin A\} .
\end{equation}

Since $\rho$ is supermodular  $\kappa \leq 1$. Moreover,
it can be shown that $k_{ij}$ is a non-increasing set function (see \cite{Hunt-Algo}) and thus its maximum occurs at sets of cardinality $m$. This increases
the practicability of the computation in (\ref{E:curvature}).

Suppose $S \subset T \in \mathnormal{L}_{\nu,C}$. Given $\nu$ we want to determine
the minimum size of $S$ for which $\rho(S) \geq \nu$. 
If $T \setminus S=\{j_1, \cdots j_{r} \}$ we have (see equation (2) \cite{Wang}) ,
\begin{equation} \label{E:setdiff}
\rho(T)-\rho(S)=\sum_{t=1}^{r}\rho_{j_{t}}(S \cup \{j_1 , \cdots j_{t-1}\}).
\end{equation}
Therefore ,
\begin{equation} \label{E:rankineq}
\rho(T)-\rho(S) \leq \rho_{j_1}(S)+\kappa\rho_{j_2}(S)+ \cdots \kappa^{t-1}\rho_{j_{r}}(S)
\end{equation}
Suppose $\rho(T)=1$, for example if $T$ is a vertex cover with $|T|=C$. Define $\gamma$ to be $\gamma=\max\{\rho_{j_{t}}(S): S \subset T, t=1 \cdots r\}$.
 We can get a lower bound on the rank of $S$ using equation (\ref{E:rankineq}) and the inequality $0 \leq \rho_{j}(S)\leq \gamma$. First assume $\gamma$ is known. We know that if $S \ne \emptyset$, then $\gamma < 1$. Then,
\begin{equation} \label{E:ranklower}
\rho(S) \geq 1-\gamma\sum_{t=1}^{r}\kappa^{t-1}
\end{equation}
Let us now suppose that :
\begin{equation}\label{E:rank&nu}
 (1- \gamma\sum_{t=1}^{r}\kappa^{t-1})\geq \nu ,
\end{equation}
and $|S|\geq m$. If an approximation of quality $\nu$ is required, and
$r(\nu)$ is the largest value of $r$ such that inequality (\ref{E:rank&nu}) holds, then $r  \leq r(\nu)$. Now $K=C-r$ is the
cardinality of $S$ so that $C-r(\nu) \leq C-r$. Thus the smallest possible value of $|S|$ is
\begin{equation}
m(\nu)=C-r(\nu)
\end{equation}
In particular any $m$ must satisfy $m \geq m(\nu)$.
Conversely, given $m$, the quality of the approximation depends on $\gamma$, the largest marginal increase of a set $S$ of size $m$,  
$\kappa$ and $r=C-m$. More precisely, the largest value of $\nu$ and thus the guaranteed quality of an approximation obtained by our method, has an upper bound given by the left hand side of  (\ref{E:rank&nu}). 

\section{Conclusion}
In a simple model of communication based on a random walk in an undirected graph, the problem of finding the subset of nodes of defined cardinality  that enable the fastest communication in the network is posed in terms of finding the target set that minimizes the sum of the first arrival times of random walkers starting outside the set. The problem is probably NP complete as stated. Thus we sought approximations based on constraining the search space to so-called optimal and near optimal sets of cardinality bounded by some  constant $C \,$, the cardinality of a vertex cover.  We defined a collection of optimal and near optimal sets of pre-defined quality $\nu$ and constrained our search for approximations of the minimization problem to these sets of cardinality no more than $K$ where
$m < K < C$, and $m$ is the minimum cardinality of sets of quality $\nu$. The offered approximate solution to the problem was then obtained by greedy extension of each member of a selected starter set of near optimal sets of cardinality $m$ 
(see section  \ref{S:optnopt} and Section \ref{S:closure}). We show that the ratio of the ranks of the our approximate solution to the exact solution is no worse than the corresponding ratio for the classic greedy solution and in general improves the ratio by a constant that can be calculated once the approximation is known. Moreover using the concept of curvature for submodular functions, we were able to quantify the tradeoff between $\nu$ , the lower bound on the quality of the approximation and the computational effort as measured by $m$.
The stated computational effort $O(N^{m+3})$ (see section \ref{S:effort})  was based on a very conservative  estimate of the computation needed to obtain the starter set. Indeed we believe incorporating more knowledge about the "graph structure" of optimal and non-optimal sets would greatly reduce this estimate. We conjecture that $m$ is quite small for graphs bipartite graphs and even graphs of finite tree dimension because the structure of the near optimal and optimal sets for these graphs do not require that the quality of the sets (as measured by $\nu$) be high. Current research on this conjecture is underway.





%
%
%

\end{document}